\documentclass[manuscript, nonacm]{acmart}

\renewcommand\footnotetextcopyrightpermission[1]{}

\AtBeginDocument{%
  }

\setcopyright{cc}
\setcctype{by}

\begin{document}

\title[``Zooming In'' on Agentic Web Browsers as Assistive Technologies]{``Zooming In'' on Agentic Web Browsers as Assistive Technologies: A Case Study with a Low-Vision Technology Expert}
\author{Laura Colazzo}
\orcid{0009-0005-0274-5991}
\affiliation{%
  \institution{Politecnico di Milano}
  \city{Milan}
  \country{Italy}
}
\email{laura.colazzo@polimi.it}

\author{Giuseppe Anzillotti}
\orcid{0009-0002-3880-6352}
\affiliation{%
  \institution{Politecnico di Milano}
  \city{Milan}
  \country{Italy}
}
\email{giuseppe.anzillotti@mail.polimi.it}


\begin{abstract}
Agentic Web Browsers (AWBs), powered by Large Language Models (LLMs), are emerging as autonomous systems capable of navigating the Web on behalf of users. Beyond enhancing productivity, they could also offer significant promise as Assistive Technologies (ATs) for visually-impaired individuals, transforming web interaction into a fluid conversational exchange. In this paper, we present a case study with a low-vision technology expert, examining how AWBs can support visually-impaired users in web navigation. The findings show that, despite the current limitations, the navigation experience is notably fluid and flexible, underscoring the strong potential of AWBs to enhance accessibility and reduce barriers in web interaction, with implications that may extend beyond accessibility to agentic UX more broadly.
\end{abstract}


\maketitle

\section{Introduction}
Large Language Models (LLMs) are evolving from passive information-retrieval tools into active agents, capable of perceiving, reasoning, and acting in digital environments ranging from individual applications to full operating systems. Agentic Web Browsers (AWBs) exemplify this shift by leveraging LLMs to autonomously navigate and interact with the Web, translating natural language intentions into interface actions. Beyond the advantages AWBs can introduce in terms of efficiency and productivity, they also hold promise as Assistive Technologies (ATs) for people with vision impairments. 
By supporting natural-language interaction with web content, they may reduce reliance on visual webpage inspection and direct interface manipulation, thereby lowering accessibility barriers.
As a result, AWBs may represent a compelling alternative to traditional ATs, transforming web interaction into a more fluid conversational experience.

In the following, we present a case study involving a low-vision technology expert exploring the use of AWBs as assistive technologies for web navigation. 
The study aimed to gather preliminary insights into the overall experience, as well as to elicit informed perspectives, grounded in the subject's technological expertise, on the potential of these systems to reshape web accessibility for visually-impaired users.

\section{Related Work}

\subparagraph{\textbf{Agentic Web Browsers.}}
Compared to Computer-Use Agents (CUAs), such as Anthropic’s Computer Use Tool, which are capable of simulating human interaction (e.g., moving a cursor, clicking buttons, scrolling) with operating systems and a broad range of applications \cite{gubbi2026a11ycua,anthropic_introducing_2024,cheng_mapping_2026}, AWBs can be viewed as a specialized subset of these systems operating exclusively within the browser environment. Similarly to CUAs, AWBs such as ChatGPT Atlas\footnote{https://chatgpt.com/atlas/}\cite{openai_introducing_2025} or Perplexity Comet\footnote{https://www.perplexity.ai/comet} are powered by LLMs and are capable of interpreting webpage content leveraging the DOM structure, as well as screenshots taken in real-time during navigation. Taking advantage of this data, they can autonomously plan and execute actions on the Web in response to natural language user instructions.
Their graphical interface typically consists of a central web content area and a side chat panel for issuing commands to the system. When control is handed over to the agent, a colored animation appears around the frame of the central content area, while interactions with UI elements, such as clicks or scrolls, are visually displayed in real time. The interface of some AWBs, such as Perplexity Comet, also includes dedicated voice mode or a text-to-speech option to enable vocal interaction.

\subparagraph{\textbf{LLMs and web accessibility.}}
Some prior studies on LLMs and web accessibility have preserved the existing AT paradigm, leveraging Generative AI  (GenAI) capabilities to reduce the effort associated with screen-reader-based navigation rather than replacing it~\cite{ghosh2024savant, gubbi2025taskmode}.
Some recent work, however, move away from this approach and instead explore CUAs as an alternative to traditional ATs for supporting visually-impaired users. 
For example, \citet{shin_toward_2026} explored the online shopping experience of a group of users with visual impairments using a voice-first LLM-powered CUA. Despite these efforts, \citet{gubbi2026a11ycua} highlight the accessibility gap of CUAs, showing that interaction with such systems remains challenging for blind and low-vision users, as they are primarily designed with sighted people in mind, making oversight and control difficult for users with visual impairments.

Building on this body of work, this paper narrows its scope to AWBs, providing a preliminary exploration of their role in web accessibility for visually-impaired people, based on the experience of a low-vision user.

\section{Methodology}
\subparagraph{\textbf{Participant.}}
The study involved a low-vision male individual aged 31 with congenital low vision. His condition is characterized by complete blindness in the right eye and residual vision in the left eye (<1/20) accompanied by a reduction in the visual field below 3\%. The subject has used ATs since early childhood, starting with his first use of a personal computer, specifically screen zoom and color inversion, while reporting no use of speech synthesis.
Additionally, the participant is a Computer Engineer and currently employed as a professional in the technology sector. He reported advanced knowledge of GenAI and general familiarity with AWBs.

\subparagraph{\textbf{Procedure.}}
The session took place at the end of May 2026 and was conducted by two researchers: one led the interaction with the participant, while the other observed and took notes. The study involved the use of the Voice User Interface (VUI) of Perplexity Comet, an AWB with which the participant had never interacted before. Inside the Comet interface, the VUI can be accessed through a dedicated button in the top-right corner of the window or via a configurable keyboard shortcut. Upon activation, an auditory cue signals that the system is ready to receive voice input, after which the user can start interacting with the assistant through voice and interrupt its responses at any time. After collecting demographic data, the participant was given time to familiarize himself with the VUI and was then observed performing tasks in two different scenarios. In the first one, he was invited to complete an exploratory task on a commercial website consisting of locating and configuring a product, while the second scenario involved a form-filling task on a public administration portal. Following tasks completion, an in-depth semi-structured interview was conducted to gather reflections on the overall experience and elicit perspectives on the potential of this technology as an AT for navigating the Web for visually-impaired individuals.

\section{Findings}
In this section, we present the strengths and limitations that emerged from the case study, focusing on the user experience and the implications they have for the future success of this technology as an assistive tool.
\subparagraph{\textbf{High conversational fluidity and interaction flexibility.}} 
The participant expressed strong appreciation for the conversational fluidity and overall flexibility of the interaction with Comet, identifying these as key strengths of the system and as clear evidence of the potential of this interaction paradigm for the future of web accessibility. In particular, he appreciated the system’s ability to adjust the level of detail in its responses according to user contextual needs.

\subparagraph{\textbf{Lack of non-visual feedback.}}
A key issue that emerged during the interaction with Comet is the lack of non-visual feedback while the system processes requests and executes actions autonomously. This became particularly critical when the participant experienced a longer delay while waiting for a system response. Although when this occurred he understood that the system was still working, thanks to the colored animation perceived through his residual vision on the screen and his knowledge of the technology---namely, that generative processes may require some time to complete---this lack of non-visual feedback creates in general transparency issues, leaving users uncertain about what is happening on screen.

\subparagraph{\textbf{Limited control and transparency.}} Another observed issue is the lack of user control during the system's autonomous execution. In particular, in the filling-form scenario, Comet autonomously inserted fabricated data into the fields without asking for confirmation. Likewise, in the other scenario, the agent failed to inform the user of an available configuration option and chose one on his behalf. These behaviors reduced the user's sense of control, compromised system transparency, and ultimately undermined trust.

Overall, the participant's optimism about the crucial role AWBs will play in the future of web accessibility emerged consistently throughout the session. In his view, the limitations encountered can be overcome by conducting further user studies and refining the underlying technology based on user needs, steering the web agent to implement interaction mechanisms that foster user control, enhance transparency and trust, and still provide a fluid interaction experience.

\section{Conclusions and Future Work}
In this paper, we presented a case study on the use of Perplexity Comet as an assistive tool for web navigation by a low-vision technology expert. 
Although findings are based on a single-case study and cannot be generalized to the broader population of visually-impaired users, they offer an initial positive perspective on the potential of AWBs as ATs. This perspective is shaped by the participant's technical expertise and his ability to recognize the system’s potential beyond its current limitations. Future work should therefore continue this line of research, extending the investigation to a larger and more diverse group of participants, including blind individuals, with particular attention to issues of transparency, user control, and trust. Comparative evaluations with other existing ATs, such as screen readers, could also be conducted to better contextualize the role of AWBs within the accessibility landscape. 

More broadly, our findings suggest that agentic web browsing can be a promising interaction paradigm, as it can reduce the effort required to navigate the Web and enable more natural forms of interaction. To fully realize this potential, however, we believe that the research community should direct its efforts to making it more inclusive. 
Indeed, we argue that the principles of inclusive design are particularly relevant in this context, and the limitations identified in this study, such as the limited user control, may arise more in general whenever users delegate tasks to agentic systems, regardless of their ability, reflecting broader challenges in agentic UX.
We, therefore, believe in the importance of approaching the design of agentic web browsing experiences through the lenses of inclusivity, recognizing that accessibility-driven improvements may generalize beyond their original target users, e.g., people with visual impairments, and benefit a broader population. 
\bibliographystyle{ACM-Reference-Format}
\bibliography{sample-base}
\end{document}